\documentclass[prb,aps,  showpacs, reprint, preprintnumbers,amsmath,amssymb,superscriptaddress,floatfix]{revtex4-1}

\usepackage[]{graphicx}      
\usepackage{bm}             	
\usepackage{times}			
\usepackage{tabularx}
\usepackage{color}
\usepackage{dcolumn}		
\usepackage{amsmath}
\usepackage{amssymb}
\usepackage{amsfonts}
\usepackage{times}
\usepackage{tabularx}
\usepackage{xcolor,colortbl}
\newcommand{\IEF}{Institut d'Electronique Fondamentale, CNRS, Univ. Paris-Sud, Universit\'e Paris-Saclay, 91405 Orsay, France}
\newcommand{\IMEC}{imec, Kapeldreef 75, B-3001 Leuven, Belgium}

\definecolor{lightgray}{gray}{0.9}

\begin{document}
\title{Time-resolved spin-torque switching in MgO-based perpendicularly magnetized tunnel junctions}

\author{T. Devolder}
\email{thibaut.devolder@u-psud.fr}
\author{Joo-Von Kim} 
\author{F. Garcia-Sanchez}
\affiliation{\IEF}
\author{J. Swerts}
\author{W. Kim}
\author{S. Couet}
\author{G. Kar}
\author{A. Furnemont}  
\affiliation{\IMEC}

\date{\today}                                           
%
%
\begin{abstract}
We study ns scale spin-torque-induced switching in perpendicularly magnetized tunnel junctions (pMTJ). Although the switching voltages match with the macrospin instability threshold, the electrical signatures of the reversal indicate the presence of domain walls in junctions of various sizes. In the antiparallel (AP) to parallel (P) switching, a nucleation phase is followed by an irreversible flow of a wall through the sample at an average velocity of 40 m/s with back and forth oscillation movements indicating a Walker propagation regime. A model with a single-wall locally responding to the spin-torque reproduces the essential dynamical signatures of the reversal. The P to AP transition has a complex dynamics with dynamical back-hopping whose probability increases with voltage. We attribute this back-hopping to the instability of the nominally fixed layers.
\end{abstract}


\maketitle

%
%

The spin-transfer-torque (STT) manipulation of the magnetization is a cornerstone of modern spintronics.  In magnetic tunnel junctions (MTJ), the interplay between magnetization-dependent transport properties \cite{theodonis_anomalous_2006, tang_influence_2010} and the spin torques results in a rich variety of phenomena \cite{locatelli_spin-torque_2014}.  After the discovery of STT, it was soon realized \cite{sun_spin-current_2000, bazaliy_towards_2001} that the cylindrical symmetry of the magnetic properties in Perpendicular Magnetic Anisotropy (PMA) systems and the resilience to thermal fluctuations that the anisotropy provides would make PMA systems ideal playgrounds to explore STT-induced dynamics. However MTJs with relevant properties  became available only a decade after \cite{ikeda_perpendicular-anisotropy_2010} and relied on ultrathin systems where strong interfacial effects can be present \cite{cho_thickness_2015}; besides, efficient spin-torque generation requires  complex embedding stacks \cite{gajek_spin_2012, swerts_beol_2015} in which each additional layers can be a fluctuator strongly coupled to the layer of main interest in a non uniform \cite{gajek_spin_2012, Bernstein_nonuniform_2011} and non local \cite{tserkovnyak_nonlocal_2005} manner. As a consequence the STT-induced magnetization switching in PMA MTJ systems exhibits rich features \cite{worledge_spin_2011, sun_effect_2011} that deserve to be studied, especially as it opens opportunities in information technologies.

In this letter, we report single-shot time-resolved measurements of ns-scale STT switching events in PMA MTJs. We detail the electrical signature of the switching and account for its main features using a simple formalism. After an observable nucleation, the reversal proceeds in a non uniform manner with the motion of a domain wall (DW) in a Walker regime; this comes together with intensified excitations in the nominally fixed layers that can result in dynamical back-hopping. This complex dynamics calls for a revisit of the models describing the stability of magnetization and its switching under STT in perpendicularly magnetized confined systems. Our findings are also important for the understanding of other spin torque devices like spin majority gates \cite{nikonov_proposal_2011} where the degree of coherence of the magnetization --the occurrence or non occurrence of domain walls-- is crucial.\\
The paper is organized as follows. We first describe in detail the properties of the thin films from which the samples are fabricated (section~\ref{FilmProp}). The device switching properties are then reported both in quasi-static limit and in a time-resolved manner (section~\ref{devices}). The reversal is modeled assuming a domain wall mediated process (sections~\ref{wallmodel1}-\ref{wallmodel2}) or dynamically active fixed layers (section~\ref{BH}).

\section{Properties of the thin film samples} \label{FilmProp}
We use bottom-pinned MTJs of the following configuration \cite{swerts_beol_2015} : seed / Hard Layer / Ru (0.85 nm) / Reference Layer / Ta (0.4 nm) / Spin Polarizing Layer / MgO (${RA_{p}}=6.5 ~\Omega.\mu \textrm{m}^2$)/ Free Layer / cap. The Free Layer (FL) is a 1.4 nm thick FeCoB layer optimized for high TMR (150 \%). The fixed layer is constructed in a synthetic ferrimagnet configuration for stray field compensation. It consists of three parts: the 1.3 nm thick FeCoB spin polarizing layer (PL) whose \textit{fixed} character is ensured by a ferromagnetic coupling with the Co-Pt based reference layer (RL) though a Ta spacer. The RL is hardened by an antiferromagnetic coupling with a thicker Co-Pt based hard layer (HL) through a Ru spacer.  

\begin{figure}
\includegraphics[width=9cm]{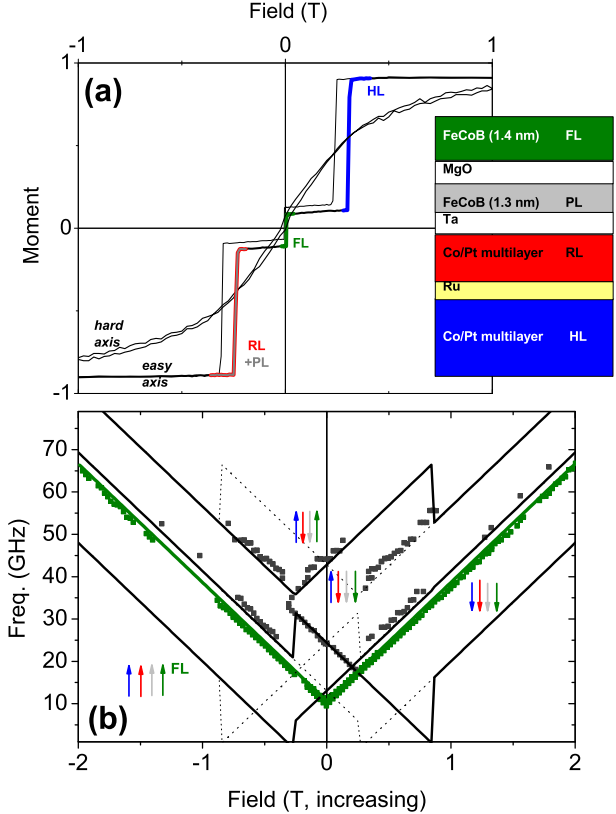}
\caption{Properties of the unpatterned MTJ. (a) Hard and easy axes loops; the easy axis recorded in decreasing field appears in a light grey color. (b) Spin wave frequencies measured (symbols) and calculated (bold lines) during an easy axis loop upon increasing field. The dotted lines are the calculated modes of the other metastable configurations undergone during a decreasing field loop. Inset: sketch of the MTJ.}
\label{wafer_level_measurements}
\end{figure}

The film easy axis loop [Fig. \ref{wafer_level_measurements}(a)] indicates that all layers have perpendicular magnetization and that they switch sequentially. As common for soft materials like FeCoB, the FL switches at a few mT. The HL switches at 0.3 T. The PL and the RL stay rigidly coupled to each other and they switch in synchrony at -0.25 T. The negative sign recalls their antiferromagnetic coupling with the HL. Hard axis loops [Fig. \ref{wafer_level_measurements}(a)] indicate that a field above 1.2~T is needed to saturate the MTJ. Hence the coercivities are extrinsic and cannot be used to quantify the properties of each layer of the MTJ. 
%

Each of the four block of the MTJ has a specific ferromagnetic resonance (FMR) signature that we could detect [Fig. \ref{wafer_level_measurements}(b)]. Some of the modes can be understood from qualitative arguments. This is the case of the V-shaped mode [Fig.~\ref{wafer_level_measurements}(b), green symbols] which bends at the free layer (FL) coercivity, and which must be assigned to the FL. The other modes are greatly affected by strong exchange through Ru and Ta such that they involve magnetization motion in all 3 parts of the fixed system. A fit to the modes of coupled macrospins was used to determine each layer's properties and their interlayer exchange coupling (Table I), with a procedure to be described elsewhere. The properties include the FL and PL damping factors of $0.01 \pm 0.001$ and $0.015 \pm 0.004$. The respective stabilities of the FL and the fixed system are given by the zero field frequencies of their eigenexcitations $\omega_{\textrm{H=0}} / \gamma_0$ of 0.38~T and 0.55~T ($\gamma_0$ is the gyromagnetic ratio).

\begin{table*}
  \centering
  \begin{tabular}{c|c|ccccl}
  \hline
Layer  & Free layer & Spin polarizing layer (PL) & & Reference Layers (RL)  &  & Hard layers (HL) 	\\ \hline  \hline
Composition  & Fe$_{60}$Co$_{20}$B$_{20}$ & Fe$_{60}$Co$_{20}$B$_{20}$ & Ta &  [Co(0.5)/Pt(0.3)]$_{\times 4}$  & Ru &[Co(0.5)/Pt(0.3)]$_{\times 6}$ /Co(0.5)\\ \hline

Thickness ($t$, nm) & 1.4 & 1.3 &  0.4 &   3.2 & 0.85 &5.3 \\ \hline
$M_S$ (A/m) $\dagger$ &  $1.1\times 10^6 $ & $ 1.1\times10^{6} $ & & $ 8\times10^{5} $  & & $8.5\times10^{5}$ \\  \hline
$\mu_0(H_{k}-M_S)$ (T)   &  $0.38\pm 0.01$ & $0.13 \pm 0.05$ & & $0.5 \pm 0.05$  & & 0.63 $\pm$ 0.05\\  \hline 
$J$ (mJ/m$^2$) $\dagger$ & -  & & $J_{\textrm{Ta}}$=~0.8 & &$J_{\textrm{Ru}}$=~-~1.5 \\  \hline 
$\alpha$  & $0.01 \pm 0.001$ & $0.015 \pm 0.004$ &   & - &   & - \\  \hline   
$\alpha \frac{2e}{\hbar} RA_{p} \eta \frac{\omega_{H=0}}{\gamma} M_S  t $ & $ \approx 0.4$~V & $0.6 \pm 0.2$~V &   & - &   & -\\[1ex]  \hline

    \end{tabular}
  \caption{Set of properties consistent with the eigenmode frequencies of the unpatterned MTJ. The symbol $\dagger$ recalls that the corresponding quantity relies on a subjective choice of effective magnetic thicknesses. The STT efficiency $\eta$ stands for $(1+p^2)/p$ where $p\approx 1$ is the spin polarization.}
  \label{bilan}
\end{table*}

Using the data of Table I, the FL macrospin instability threshold should be around 0.4~V. We emphasize that 0.4 V should yield an instability of the FL uniform state state, but this does not imply that the reversal happens in a uniform manner. While this low threshold is promising in the sense that it is far below the voltages leading to material degradation, two points are to be noticed. \\
(i) The areal moments of HL vs \{RL + PL\} are too imbalanced for a perfect stray field compensation, which may affect the two switching transitions differently \cite{gopman_bimodal_2014}. Magnetostatics was used to predict their vector stray field H (Fig.~\ref{DUT07loop}, inset). In the central part of the FL, $H_z$ has a plateau favoring the AP orientation while near the edges it favors P. Experimentally a P coupling is seen, consistent with a field-induced reversal though a nucleation at some edge. Note that the stray field is along the (z) axis only at the center of the FL: everywhere else there is a substantial in-plane component $H_x$.
\\
(ii) The coupling through Ta ensures that the PL and RL magnetizations are parallel at remanence. However, the moderate stiffness field of the PL and its low thickness $t$ are such that its foreseen macrospin instability threshold $\alpha \frac{2e}{\hbar} RA_{p} \eta \frac{\omega_{\textrm{H=0}}}{\gamma} M_S  t $ is only $0.6 \pm 0.2$~V. This is only slightly above that of the FL (0.4 V, Table I). One may thus question to what extent the RL magnetization can stay static when large voltages are applied.

\section{Properties of the patterned devices} \label{devices}

\subsection{Quasi-static switching} 
\begin{figure}
\includegraphics[width=8cm]{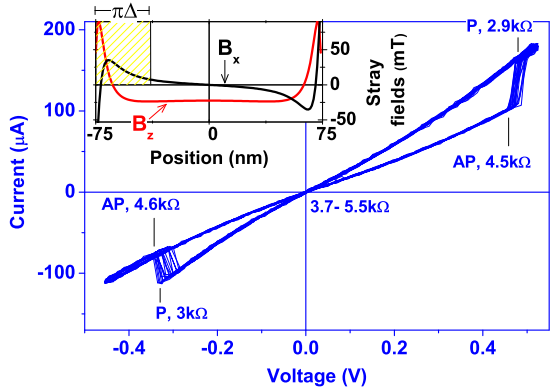}
\caption{(Color online) 16 current-voltage loops of a 75$\times$150 nm$^2$ MTJ when compensating the stray field of the fixed layers. Inset: Length cut of the in-plane ($B_x$) and out-of-plane ($B_z$) stray fields at the FL position. The shaded area denotes the DW width $\pi \Delta$.}
\label{DUT07loop}
\end{figure}

The MTJs were patterned into pillars of various shapes from rounded rectangles of $75\times150$ nm$^2$ to circles of 500 nm diameter. Minor loops of the FL indicate some size-dependent coupling with the fixed system. The magnetostatics calculations Fig.~\ref{DUT07loop}, inset) indicate that this coupling comes from the fixed layers stray field which comprises also in-plane components.  In practice, an external (uniform) out-of-plane field $H_z$ is used to empirically compensate for the (non uniform) stray field to get a centered STT loop [Fig.~\ref{DUT07loop}] where $V_{P \rightarrow AP} \approx - V_{AP \rightarrow P}$. The switching voltage is then $\approx0.4~\textrm{V}$, in line with the expectations of when the single-domain state should be destabilized. Noticeably, the distribution of $V_{AP \rightarrow P}$ is substantially narrower than that of $V_{P \rightarrow AP}$. Let us time-resolve the switching to understand this difference. 

%

\subsection{Time resolved dynamics}

\subsubsection{AP to P switching and domain wall process}
\begin{figure*}
\includegraphics[width=16 cm]{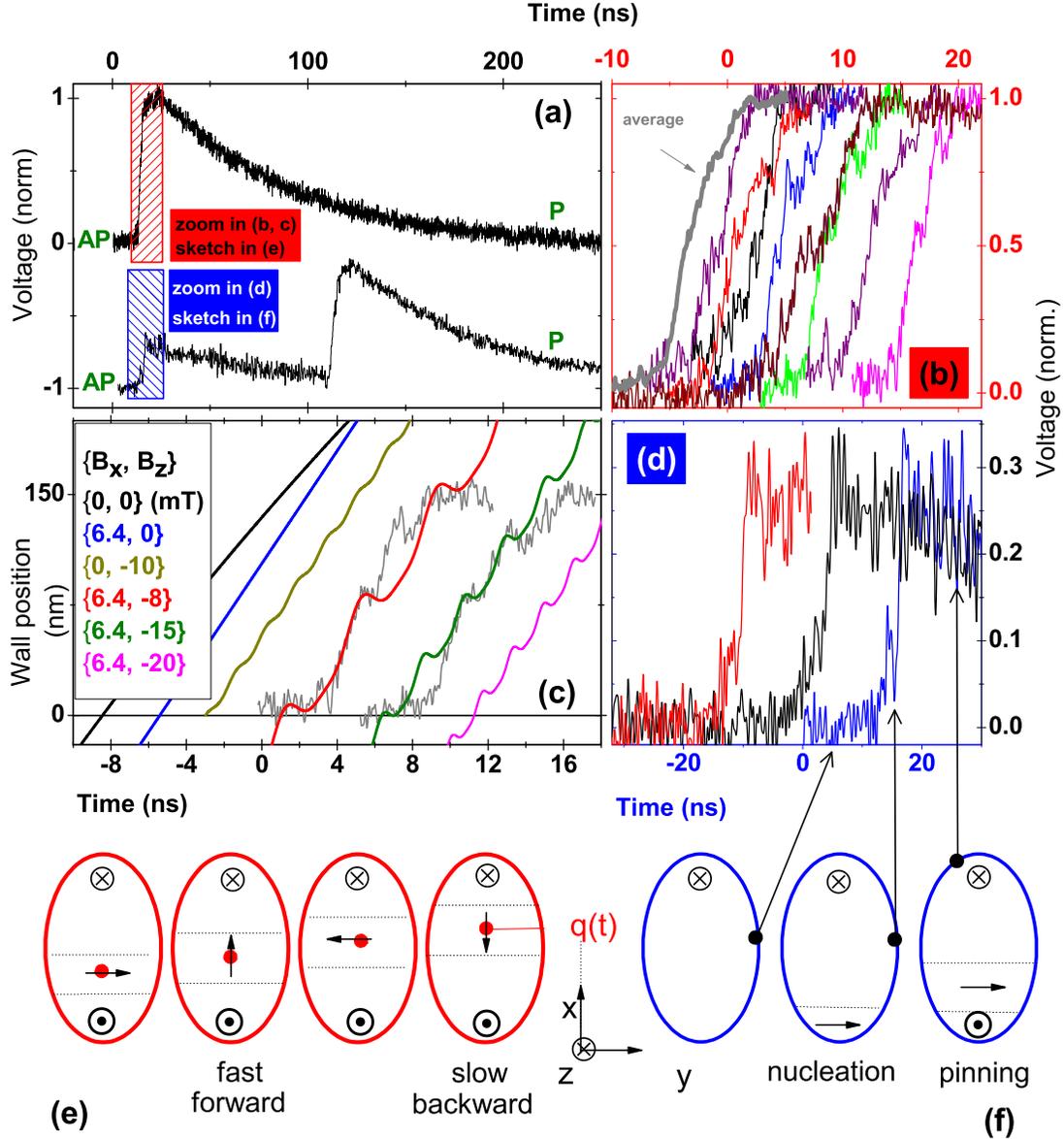}
\caption{ (Color online) AP to P switching of a 75$\times$150 nm$^2$ device.  (a) Selected switching events representative of the two class of switching events. The exponential decays ($\tau=68$ ns) after the voltage steps are capacitive artefacts related to the experimental set-up. The unambiguous labeling of the states (AP or P) is done thanks a circuit derivation (see suppl. document). (b) Zoom on 8 single-step switching events, and average thereof for a triggering criterion at half of the switching signal. The events are been time-delayed for clarity. (c) DW positions as simulated in the 1D model for  $\sigma j= 1.7$~GHz and for various field conditions mimicking the dipolar coupling with the fixed layers. The experimental (noisy) traces have been converted to DW positions assuming a single wall sweeping though the length of the MTJ. (d) Zoom on the onset of two-step switching events. (e) Sketch of the single-step reversal scenario simulated in panel (c). The dotted lines denotes the wall width and the arrow the magnetization tilt $\phi$ within the DW at position $q$. (f) Sketch of the nucleation scenario at the onset of the reversal.}
\label{DUT07APP}
\end{figure*}

\begin{figure}
\includegraphics[width=8 cm]{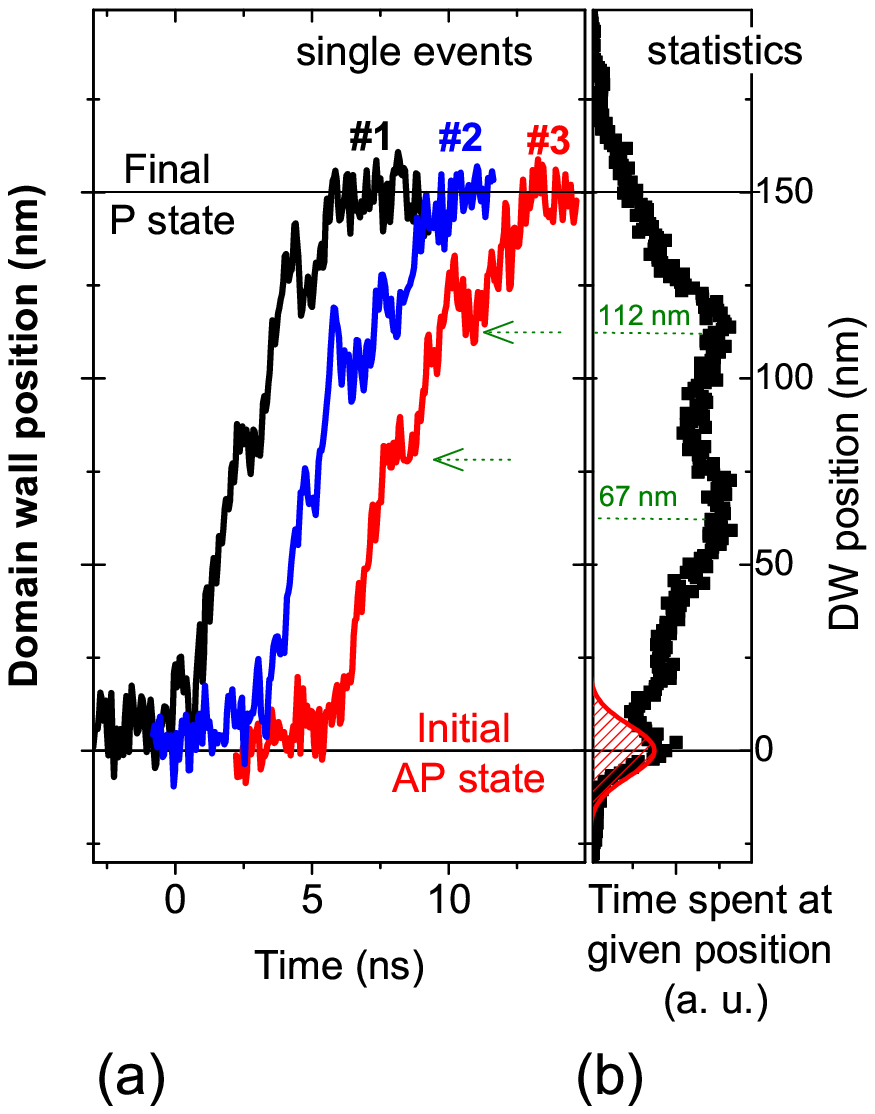}
\caption{ (Color online) Statistical analysis of the AP to P transitions of a 75$\times$150 nm$^2$ device.  (a) Examples of single-step switching events. The dashed arrows indicate the positions at which the domain wall velocity seems to vanish or transiently reverse. (b) Histogram of occurrence of a given wall position during a statistical set of single-step switching events. The red Gaussian distribution sketches an estimate of the experimental noise in the determination of the wall position as due to voltage noise. The dashed lines at the two maxima of the histogram indicate the domain wall positions at which the wall velocities are most likely to vanish.}
\label{histo}
\end{figure}

We first focus on the AP to P switching in $75\times 150~\textrm{nm}^2$ devices. Fig.~\ref{DUT07APP} displays single-shot switching curves that are representative of the AP to P switching for all the studied devices.  Irreversible and complete switching is systematically observed for this transition but the events can be casted in two categories. Since we will see that the reversal proceeds through a DW process, it is convenient to translate the voltages $v(t)$ into a domain wall position by defining the length $q$ by $$q/L=v(t)/v_\textrm{max}$$ where $v_\textrm{max}$ is the voltage change for full switching and $L$ is the device length. \\
(i) Most of the time, the AP to P switching results in a single-step ramp-like evolution of the device resistance, with a switching in 3-4 ns [Fig.~\ref{DUT07APP}(a-b)]. The fine structure of the switching events [Fig.~\ref{DUT07APP}(b, d)] indicates that the overall dynamics slows down, or makes apparent sub-ns pauses at intermediate resistance levels (intermediate DW positions). Although the DW positions at the pauses vary from event to event [Fig.~\ref{histo}(a)], they occur with maxima of probabilities at definite positions, which are $q=67\pm7$ and $q=112\pm7~\textrm{nm}$ for $75\times 150~\textrm{nm}^2$ devices [Fig.~\ref{histo}(b)]. \\
 (ii) In the other cases [Fig.~\ref{DUT07APP}(a), lower trace], the AP to P switching proceeds in two steps: a first resistance change equivalent to $q=45~\textrm{nm}$, followed by a transient pause whose duration varies stochastically in the 100 ns range, and finally a second signal rise with one or two sub-ns pauses (not shown) on top of a ramp-like evolution of the resistance in 2-3 ns.  The fine structure of the onset of switching [Fig.~\ref{DUT07APP}(d)] indicates that the reversal starts by a gradual evolution of the resistance till $q=15 \pm 3~\textrm{nm}$, lasting the initial 2-3 ns. Then the resistance ramps in 500 ps either to the intermediate resistance level at  $q=45~\textrm{nm}$ or till saturation. 
  
%

These electrical signatures of the AP to P switching cannot be understood in the framework of the macrospin approximation in which pre-oscillations and post-oscillations would be seen \cite{sun_spin-current_2000, zhang_micromagnetic_2010}. The plateau in the 2-step switching is rather indicative of a DW process. Let us see the consequences if we conjecture that there is a single DW moving along the length of the pillar. 
In that case, the initial phase is a \textit{nucleation} in which a DW enters from the edge; this nucleation phase logically ends when the DW position (experimentally: 15 nm) exceeds its half width [Fig.~\ref{DUT07APP}(f)]. This sounds reasonable since our DW width $\pi \Delta$ is expected to be 34 nm ($\Delta=\sqrt{A/K}$ is the usual Bloch wall parameter).  The intermediate state may correspond to a DW \textit{pinning} event, with the inherent stochasticity of the depining process. The total reversal time indicates that the DW has an average velocity of 40 m/s.\\

 \subsubsection{P to AP switching and dynamical back-hopping}
\begin{figure}
\includegraphics[width=9.4cm]{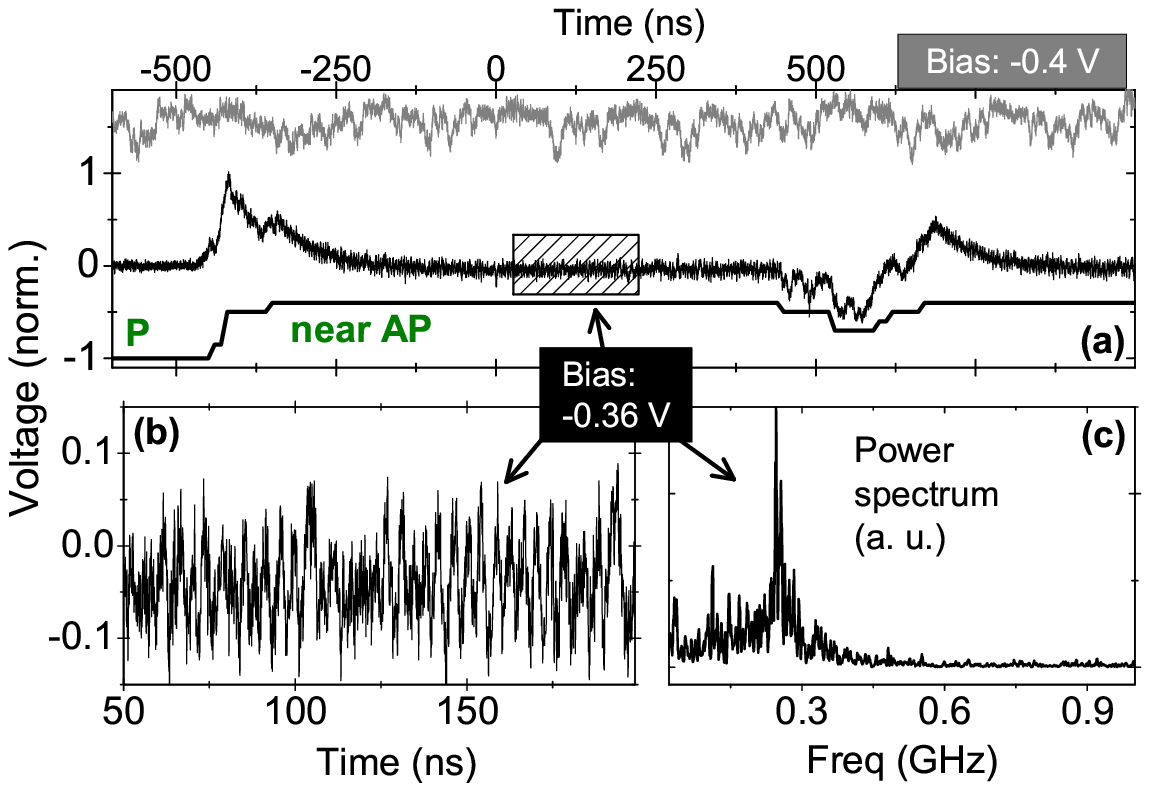}
\caption{ (Color online) P to AP switching for a 100$\times$300 nm$^2$ device.  (a) Time evolution on the signal (noisy curves) for biases of -0.4 V and -0.36 V and sketch (bold line) of the time evolution of the resistance in the latter case. The exponential decays after the voltage steps are capacitive artefacts. The labeling of the states (near AP and P) is done thanks a circuit derivation (suppl. document). (b) Zoom on the -0.36 V trace and (c) Power spectrum thereof.}
\label{DUT14PAPbh}
\end{figure}

Let us examine the reverse transition, i.e. P to AP which reveals a more complicated dynamics. It proceeds by switching attempts, incomplete saturation and dynamical back-hopping events. The events in Fig.~\ref{DUT14PAPbh} illustrate the main phenomena.
At the lowest bias inducing P to AP reversal (-0.36 V), the evolution starts by slow (20-50 ns) and gradual resistance increases to pinning levels [Fig.~\ref{DUT14PAPbh}(a)], where the systems make pauses of random durations. It then increases to a resistance level close to that of the AP state, giving an impression of full switching. However, a closer look at this state [Fig.~\ref{DUT14PAPbh}(b, c)] indicates that its resistance fluctuates. Depending on the device size and intermediate resistance level, this oscillation frequency varies from 0.2 to 2 GHz  (Supplementary Material). We have not been able to see correlations between the device size and resistance oscillation frequency. If this oscillation corresponded to a back-and-forth motion of a DW, the motion amplitude would be of 50$\pm 7$ nm for Fig.~\ref{DUT14PAPbh}(b).  These oscillatory near-AP state can survive for durations exceeding sometimes 10 $\mu$s. During this period the resistance can transiently drop [Fig.~\ref{DUT14PAPbh}(a)] in telegraph-noise manner. 

This dynamical back-hopping prevents a deterministic voltage-pulse induced STT switching. Indeed, if the voltage is switched off while the device is in such unsaturated state, the device may relax to P instead of the targeted AP state.   
One may think that a larger voltage would prevent back-hopping and force the device into the AP state; surprisingly it does the contrary (Fig.~\ref{DUT14PAPbh}). The device resistance is more and more agitated and the probability of back hopping after a current pulse increases (not shown). This increase of the rate of occurrence of back-hopping-to-P phenomena with the voltage is systematic for our sample series. We conclude that it is related to the stack properties.

%

 \section{Modeling of device switching}
 
 \subsection{Assumptions for the domain wall mediated switching}
\label{wallmodel1}
To model the AP to P switching, we assume that there is a straight DW in the system which lies at a variable position $q$ along the length axis $x$ [Fig.~\ref{DUT07APP}(e)]. Following the predictions of \cite{munira_calculation_2015} we conjecture that the DW is nucleated from the edge when the single-domain state is destabilized [Fig.~\ref{DUT07APP}(f)]. The experimental configuration (Supplemental Material) is such that the current density $j$ at the DW can be considered as constant during the switching. Using the so-called 1D model \cite{malozemoff_magnetic_1979, thiaville_micromagnetic_2005} we describe the DW as a rigid object with a tilt $\phi$ of its magnetization in the device plane (by convention $\phi=0$ for a wall magnetization along $x$), subjected to the fields $H_x$ and $H_z$. These fields are assumed to vary slowly at the scale of the DW width, which restricts the validity of our model to the sole inner part of the device (see inset in Fig.~1).  $j$ is assumed to transfer one spin per electron to the DW by a pure Slonczewski-like STT. Omitting the FL subscript, the DW can be described \cite{cucchiara_domain_2012} by the two differential equations: 

\begin{equation}
 \dot{\phi} + \frac {\alpha}{\Delta} \dot{q} = \gamma_0 H_z ~,
  \label{1Dmodela} 
  \end{equation}
  
  \begin{equation}
  \frac {\dot{q}}{\Delta} - \alpha \dot{\phi} 
 = \sigma j + \frac{\gamma_0 H_{\textrm{DW}} }{2} \sin(2 \phi) - {\frac{\pi}{2}}\gamma_0 H_x \sin \phi ~.
  \label{1Dmodelb} 
 \end{equation}
  
We have defined $\sigma = \frac{\hbar}{2e} \frac{\gamma_0}{\mu_0 M_S t}$, such that $\sigma j $ is a frequency. In our case 300 mV corresponds to 2 GHz.   
A wall parameter $\Delta=11~\textrm{nm}$ is assumed. The DW stiffness field is \cite{mougin_domain_2007}  $H_{\textrm{DW}} = (N_y-N_x) (M_S / 2)$ where $N_x \approx t / (t+w)$ is the DW in-plane demagnetizing factor when it is in a Bloch state (i.e. wall magnetization along the width $w$) and $N_y \approx t / (t+\pi \Delta)$  when it is in a N\'eel state (i. e. $\phi=0$). The DW stiffness field is for instance 10 mT when at the middle of 75~nm wide devices. It can add or subtract from the position dependent stray field $H_x$. In elliptical devices, the $H_{\textrm{DW}}$ depends on the DW position. However we will see that  $H_{\textrm{DW}}$ is not the main determinant of the dynamics, so we can consider it as constant as in infinitely long wires. 
In the absence of stray field and current, a width of $w=75$~nm yields a Walker field of $\mu_0 H_{\textrm{Walker}} = \alpha \mu_0 H_{\textrm{DW}} /2  \approx 0.05$~mT. This very small Walker field --typical of the low damped materials required for STT switching-- has implications: DW are bound to move in the Walker regime and to make the back-and-forth oscillatory movements that are inherent to this regime. While it advances on average, the DW oscillates at the frequency \cite{thiaville_domain-wall_2006} 
 \begin{equation}
 \omega_{\textrm{osc}} = \frac{\gamma_0}{1+\alpha^2} \sqrt{H_z^2-H_{\textrm{Walker}}^2} \approx {\gamma_0} |H_z|~,
 \label{DWeigenfreq} 
 \end{equation}
whose order of magnitude matches that of the observed resistance oscillations.

\begin{table*}
  \centering
\begin{tabular}{|c|c|c|c|c|} 
 \hline
   Electron direction & FL state & Torque from RL & Torque from FL & Experimental finding \\ 
     & & Effect on PL & Effect on PL &  \\ \hline
FL to PL  & P (before FL switching) & \textbf{destabilizing} PL & stabilizing PL &  \\ 
 \cline{2-4} 
  (favoring AP) &   AP (after FL switching) &  \textbf{destabilizing} PL & \textbf{destabilizing} PL & No saturation + back-hopping  \\ \hline \hline
PL to FL  & AP (before FL switching) & stabilizing PL & stabilizing  PL &  \\ 
\cline{2-4} 
(favoring P) &   P  (after FL switching) &  stabilizing PL &  \textbf{destabilizing} PL  & irreversible AP $\rightarrow$ P \\ \hline
\end{tabular}
 \caption{Torques acting on the polarizing layer before and after free layer switching. Comparison with the experimentally observed behaviors.}
  \label{backhopping}
\end{table*}

\subsection{Switching using a single wall at constant current density}
\label{wallmodel2}
Let us see the effect of current on the DW dynamics. Solving Eq.~\ref{1Dmodela} and \ref{1Dmodelb}, we find that the Walker regime is maintained for $j \neq 0$ and for $H_x \neq 0$; examples of DW trajectories are reported in Fig.~\ref{DUT07APP}(c). Two points are worth noticing: \\
The time-averaged DW velocity is changed linearly by the current.  When in the Walker regime, the current effect can be understood from Eq.~\ref{1Dmodelb}. Indeed the $\sin(2\phi)$ term essentially averages out in a time integration and the term $\alpha \dot{\phi}$ is small, such that the time-averaged wall velocity reduces essentially to: 
\begin{equation}
 \langle \dot{q} \rangle \approx   \Delta \sigma j \label{averagevelocity}
 \end{equation}
Since $\Delta=11$~nm, Eq.~\ref{averagevelocity} makes it evident that the applied voltages, which lead to $\sigma j$ in the range of a few GHz, can yield  velocities in the range of a few 10 nm/ns (i.e. in the range of a few 10 m/s).
The wall velocity is also increased by the in-plane field $|H_x|$ [Fig.~\ref{DUT07APP}(c)]. Qualitatively, this is because the wall thus stays more time with a tilt $\phi$ that maximizes its instantaneous velocity. \\
Conversely, the DW oscillation (Eq.~\ref{DWeigenfreq}) is much less affected by the current: $<\dot{\phi}>$ (Eq.~\ref{1Dmodela}) involves essentially only the out-of-plane field$H_z$. As a result, the back-and-forth DW displacement $D_{\textrm{osc}}$ is also almost not affected by the current. Indeed the terms that modulate the DW velocity are the sinus terms in Eq.~\ref{1Dmodelb}. Neglecting $\alpha^2$ and approximating  $\dot \phi$ by $\omega_{\textrm{osc}}$, the oscillatory part of the DW velocity at $H_x=0$ can be time-integrated to yield 
\begin{equation}
\frac{D_{\textrm{osc}}}{\Delta} \approx \frac{H_{\textrm{DW}}}{\sqrt {H_z^2-H_{\textrm{Walker}}^2}}
\label{Dosci}
\end{equation}
Solving Eq.~\ref{1Dmodela}-~\ref{1Dmodelb} numerically for $H_x=0$ confirms this picture. Comparing the distance covered by the DW thanks to its time-averaged (forward) velocity (Eq.~\ref{averagevelocity}) with the backward motion (Eq.~\ref{Dosci}) due to its back-and-forth (Walker-like) motion, we find that the wall only advances (it \textit{flows} forward with no transient backward motion) provided 
\begin{equation}
 \sigma j \geq \gamma_0 H_{\textrm{DW}} / \pi
 \end{equation}
 If $H_x$ were vanishing, this DW forward-only flow condition would be verified in practice. However, the fixed layers generate some in-plane stray field (Fig.~\ref{DUT07loop}, inset). The addition of $H_x$ modulates the oscillation every second period of the DW oscillation and increases the oscillation amplitude [Fig.~\ref{DUT07APP}(c)].

The comparison drawn in Fig.~\ref{DUT07APP}(c) indicates that the AP to P transition can be essentially described within this model: once nucleated at the instability of the uniformly magnetized state, the DW flows in a Walker regime with the associated accelerations-decelerations giving the impression of sub-ns pauses in the voltage traces. While moving in a region of non uniform stray field (e.g. near the device edge) the velocity is modulated by the stray field so that the distance covered between two successive accelerations varies, recalling the experimental behavior [Fig.~\ref{DUT07APP}(b)].
All together, once engaged the switching duration in a DW scenario varies simply with the inverse current: 
\begin{equation}
\tau_{\textrm{switch}} =  \frac{L}{\Delta \sigma j}
\label{tauswitch} 
\end{equation}

\subsection{Discussion on the incomplete P to AP switching and the dynamical back-hopping}
\label{BH}

The P to AP transition does not happen with the simple scenario sketched in Fig. 2(e) and described in the two previous sections (¤\ref{wallmodel1} and ¤\ref{wallmodel2}). Though the P to AP transition exhibits oscillatory features recalling the Walker regime, there are additional step-like transitions with dynamical back hopping whose probability of occurrence increases at larger applied voltages. These step-like transitions resemble more telegraph noise phenomena that oscillatory phenomena. We suspect that the step-like transitions are telegraph noise changes in the magnetic configuration of the PL. The main difference between the PL and the FL is that the PL experiences STT coming from its \textit{two} surrounding layers (FL and RL). Considering its moderate stiffness (Table I), it is conceivable that the PL be affected by STT. \\ Let us qualitatively analyze its stability.

Table 2 gathers the different spin torques acting on the polarizing layer and their expected consequence for the stability of the PL, assumed to be macrospin-like.  Let us distinguish the two current polarities.\\
(i) When the current polarity was favoring the P state (previous sections), the torques originating from the RL and FL had both PL stabilizing effects before the FL switching, and competing effects on the PL magnetization after the FL switching. In the worst case, the torques originating from the RL and FL were of opposite sign on the PL such that they could partially cancel out. In this situation, a nice irreversible AP to P transition was observed experimentally, with no indication of dynamics in the PL. \\
(ii) Conversely the torques originating from the RL and FL have a combined destabilizing effect on the PL magnetization when the current stays on after a P to AP transition (Table 2). 
This corresponds precisely to the experimental configuration in which the MTJ is found not to properly saturate in the switched (AP) state and to undergo dynamical back hopping. In this P to AP transition, the torques increasingly destabilize the PL as the FL switching proceeds. We believe that once engaged, a P to AP switching attempt might not terminate because of this feedback. While this is the likely origin of the very different reversal paths of the two transitions, the modeling of this technical problem is beyond the scope of this present study.

\section{Conclusion}
Our conclusion is twofold. Technology-wise, downsizing the junction will accelerate the reversal as long as a wall process is involved, but the mitigation of back hopping calls for a hardening of the polarizing section of the MTJ. More fundamentally, the complexity of the switching --non uniform and non symmetrical--  calls for a revisit of the nature of the STT-induced instability. It remains an open question whether the additional fluctuators present in a real MTJ comprising several magnetic layers intrinsically prevent a symmetrical bidirectional switching. Besides, the non uniform nature of the magnetic response is still present at the nanoscale; this has implications for the understanding of the numerous systems where spatial coherence of the spin system is crucial.



%

\end{document}